\documentclass[11pt]{article}
\usepackage{moriond,epsfig}

\def\be{\begin{equation}}
\def\ee{\end{equation}}
\def\bea{\begin{eqnarray}}
\def\eea{\end{eqnarray}}

\begin{document}
\begin{flushright}
LYCEN-2005-17 \\
\today
\end{flushright}
\vspace*{4cm}
\title{$K \to \pi \nu {\bar \nu}$ AND $R$-PARITY VIOLATING SUPERSYMMETRY}

\author{ J.WELZEL }

\address{Institut de Physique Nucleaire de Lyon, 4 rue Enrico Fermi, Campus La
Doua\\
69622 Villeurbanne cedex, France}

\maketitle\abstracts{Motivated by the
recent experimental data of the E787, E949 and E865 collaborations and by
the difference between the standard model ($SM$) prediction and data, we
consider in detail R-parity violating ($RPV$) supersymmetric contributions to 
$K \to \pi \nu {\bar \nu}$. The theoretical cleanness of this decay constitutes a 
useful way to provide constraints, independent of long distance effects. 
Including the possibility of interferences between one-loop $R$-parity conserving
($RPC$) supersymmetry and tree-level $RPV$ 
supersymmetric contributions, our results allow to improve the limits on 
$R$-parity violating couplings with respect to previous analyses.}

%%%%%%%%%%%%%%%%%%%%%%%%%%%%%%%%%%%%%%%%%%%%%%%%%%%%%%%%%%%%%%%%%%%%%%%%%%%%%%%%
%%%%%%%%%%%%%%%%%%%%%%%%%%%%%%%%%%%%%%%%%%%%%%%%%%%%%%%%%%%%%%%%%%%%%%%%%%%%%%%%
\section{$K \to \pi \nu {\bar \nu}$ in the standard model}%%%%%%%%%%%%%%%%%%%%%%
%%%%%%%%%%%%%%%%%%%%%%%%%%%%%%%%%%%%%%%%%%%%%%%%%%%%%%%%%%%%%%%%%%%%%%%%%%%%%%%%
%%%%%%%%%%%%%%%%%%%%%%%%%%%%%%%%%%%%%%%%%%%%%%%%%%%%%%%%%%%%%%%%%%%%%%%%%%%%%%%%
The process $K^+ \to \pi^+\nu\bar{\nu}$ is
governed in the $SM$ by the following effective Hamiltonian~\cite{BSU}
%%%%%%%%%%%%%%%%%
\begin{equation} 
H_{eff}= \frac{G_f}{\sqrt{2}} \frac{2 \alpha_e}{\pi \sin^2\theta_w} 
\sum_l \left( \lambda_c X^l_c +\lambda_t
X_t \right) \bar{s_L}\gamma^{\mu} d_L\;
\bar{\nu^l_L}\gamma_{\mu}\nu^l_L + h.c., 
\label{hamiltonian}
\end{equation}
%%%%%%%%%%%%%%%%%
where $\lambda_i=V^*_{is} V_{id}$ are products of CKM~\cite{CKM} matrix elements. The
loop-function $X_t$ contains the top contribution, and $X_c^l$ the
charm contribution for flavour $l$. In the computation of the branching ratio, 
the hadronic matrix element can be related via
isospin to the experimentally well known decay $K^+ \to \pi^0 e^+ \nu_e$
~\cite{MP}. It's branching ratio has recently been measured with high 
statistics by the E865 collaboration~\cite{Ke3}. However, their result, 
$BR(K^+ \to \pi^0 e^+ \nu_e)=(5.13\pm 0.15) \times 10^{-2}$, 
differs considerably from the most recent value of the Particle Data
Group~\cite{PDG04}, $(4.87\pm 0.06) \times 10^{-2} $,
which does not include yet the above mentioned result. So, we will use for our 
analysis an average value, where we take into account the Particle Data Group
fit as well as the E865 result: 
$BR(K^+ \to \pi^0 e^+ \nu_e)_{av}=(5.08\pm 0.13) \times 10^{-2} $.
With updated values of CKM elements and quark masses, our standard
model prediction at one-loop for the branching ratio of $K^+ \to \pi^+ \nu {\bar
\nu}$ is :
%%%%%%%%%%%%%%%%% 
\begin{equation}
BR(K^+ \to \pi^+ \nu {\bar
\nu})^{SM}=(8.2 \pm 1.2)\times 10^{-11}
\end{equation}
%%%%%%%%%%%%%%%%%
which is still compatible with the recent
experimental result~\cite{KPINUNU}, $(1.47\
^{+1.3}_{-0.8})\ 10^{-10}$. However, the predicted central
value is half the observed value so possible new physics
effects should be of the same order as the $SM$ ones in order to get the
measured central value. In this paper, we will be concerned with obtaining 
limits on $R$-parity violating couplings of supersymmetric extensions of the
$SM$. 

%%%%%%%%%%%%%%%%%%%%%%%%%%%%%%%%%%%%%%%%%%%%%%%%%%%%%%%%%%%%%%%%%%%%%%%%%%%%%%%%
%%%%%%%%%%%%%%%%%%%%%%%%%%%%%%%%%%%%%%%%%%%%%%%%%%%%%%%%%%%%%%%%%%%%%%%%%%%%%%%%
\section{$R$-parity conserving supersymmetry}%%%%%%%%%%%%%%%%%%%%%%%%%%%%%%%%%%%
%%%%%%%%%%%%%%%%%%%%%%%%%%%%%%%%%%%%%%%%%%%%%%%%%%%%%%%%%%%%%%%%%%%%%%%%%%%%%%%%
%%%%%%%%%%%%%%%%%%%%%%%%%%%%%%%%%%%%%%%%%%%%%%%%%%%%%%%%%%%%%%%%%%%%%%%%%%%%%%%%
At this stage we will assume unbroken $R$-parity. Then, just as in the $SM$, 
there are no supersymmetric contributions at tree level. They start only at 
one-loop order. 

The determination of the supersymmetric contribution is obtained in the
same way as in the $SM$ case : with the effective Hamiltonian of 
Eq.~\ref{hamiltonian} where $X_t$ is now replaced by~\cite{BRS} 
$X_{new}=r_K e^{-i\theta_K}X_t$. $r_K$ and $\theta_K$ parameterize 
new physics contributions and are functions of masses and couplings of the new 
particles. The $SM$ is then included as a special case, where $r_K=1$ and 
$\theta_K=0$. With a $MSSM$-like field content, the standard model
particles, the charged higgses, the charginos and the neutralinos enter in the
loops. 
Unfortunately the 
results for the branching ratio are very sensitive to
the yet-unknown $SUSY$ parameters (masses and couplings). However, it is possible to estimate the order of
magnitude of the $R$-parity conserving ($RPC$) contributions. The authors of Ref.~\cite{BRS} found for
$r_K$ and $\theta_K$ the typical ranges~\footnote{These ranges only indicate 
the most probable values. Our
updated analysis agrees with this statement slightly enhancing the
probability.} 
%%%%%%%%%%%%%%%%%
\begin{equation}
\label{rangeK} 
0.5<r_K<1.3,\ -25^o<\theta_K<25^o 
\end{equation}
%%%%%%%%%%%%%%%%%
by varying all $SUSY$ parameters within the bounds allowed by
experimental constraints. Then, varying $r_K$ and $\theta_K$ within 
these ranges makes at most a
change of $\sim$50\%~\footnote{However, some particular points outside the
ranges~(\ref{rangeK}) and some points of the 
parameter space of the general MSSM, where some assumptions made here have been 
relaxed~\cite{BEJR}, can give larger branching ratio and can saturate the 
experimental central value.} for the branching ratio of $K^+\to\pi^+\nu\bar{\nu}$ :
%%%%%%%%%%%%%%%%%
\begin{equation}
BR(K^+ \to \pi^+ \nu \bar{\nu})^{RPC\ SUSY}\simeq (8.2\ ^{+4.3}_{-5.2})
\times 10^{-11}~. 
\end{equation}
%%%%%%%%%%%%%%%%%
So we see explicitly that contributions of $RPC$ supersymmetry 
can be of the same order of magnitude as the standard model ones.
%%%%%%%%%%%%%%%%%%%%%%%%%%%%%%%%%%%%%%%%%%%%%%%%%%%%%%%%%%%%%%%%%%%%%%%%%%%%%%%%
%%%%%%%%%%%%%%%%%%%%%%%%%%%%%%%%%%%%%%%%%%%%%%%%%%%%%%%%%%%%%%%%%%%%%%%%%%%%%%%%
\section{$R$-parity violating supersymmetry}%%%%%%%%%%%%%%%%%%%%%%%%%%%%%%%%%%%%%%
\begin{figure}[ht]
\begin{center}
\epsfig{file=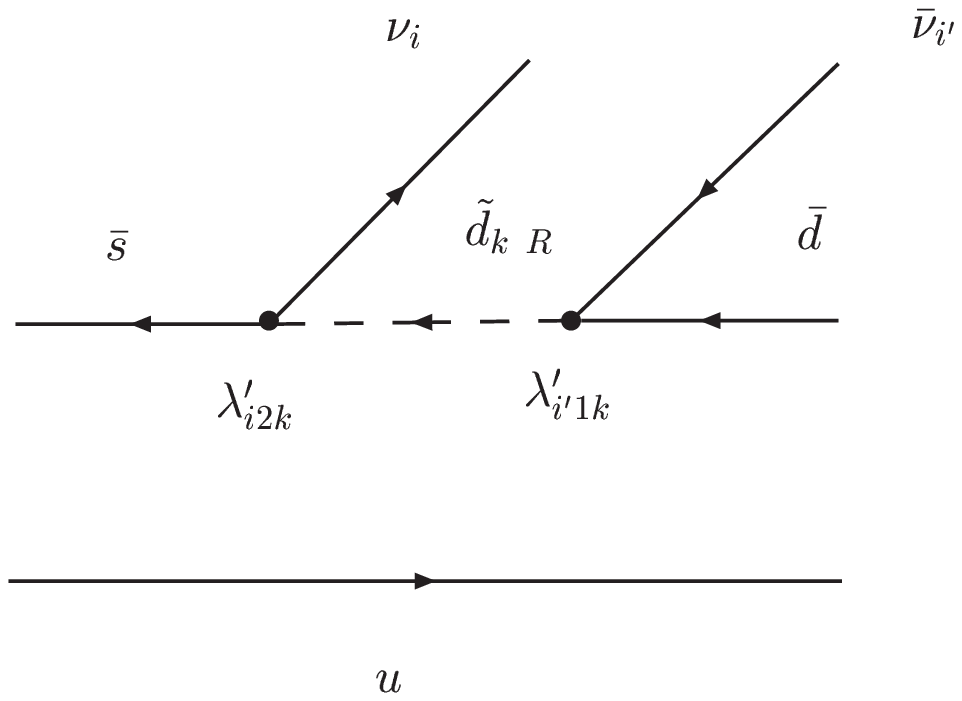,width=0.4\textwidth}
\epsfig{file=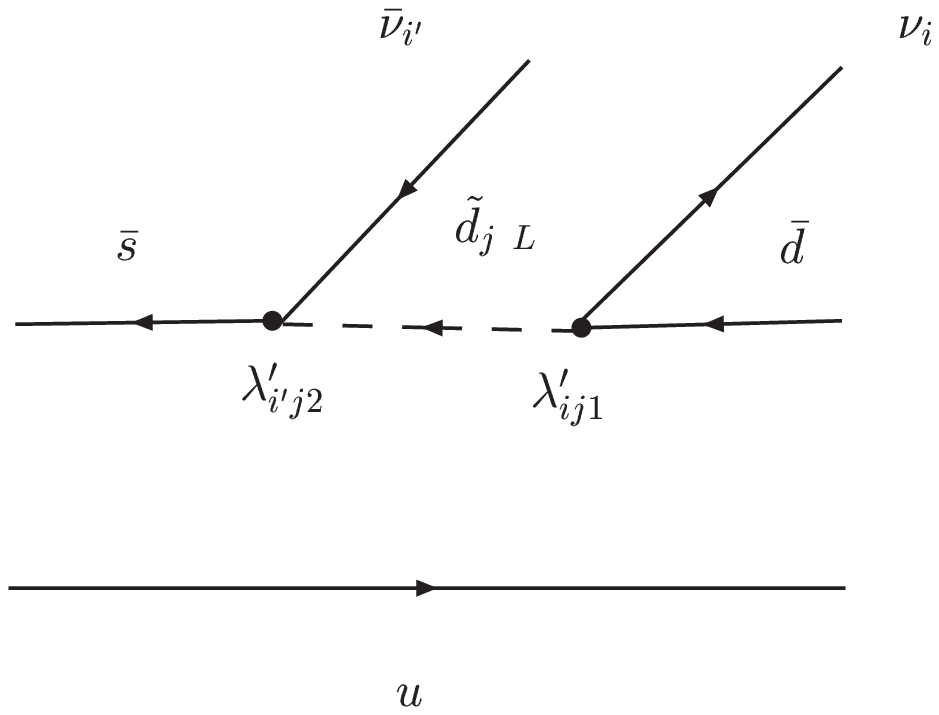,width=0.4\textwidth}
\caption{\label{fig1}\it R-parity violating tree level diagrams contributing to 
the process $K^+ \to \pi^+ \nu{\bar{\nu}}$.}
\end{center}
\end{figure}

%%%%%%%%%%%%%%%%%%%%%%%%%%%%%%%%%%%%%%%%%%%%%%%%%%%%%%%%%%%%%%%%%%%%%%%%%%%%%%%%
%%%%%%%%%%%%%%%%%%%%%%%%%%%%%%%%%%%%%%%%%%%%%%%%%%%%%%%%%%%%%%%%%%%%%%%%%%%%%%%%
By including $R$-parity violation ($RPV$) 
in a supersymmetric extension of the standard model, we have now to consider new
terms in the superpotential which allow baryon and lepton numbers violation.
These are of the form~\cite{RPV} :
\begin{equation}
W_{RPV} = \lambda_{ijk} L_i L_j E_k + \lambda'_{ijk} L_i Q_j D^c_k +
\lambda^{''}_{ijk} U^c_i D^c_j D^c_k
\label{superpotential} 
\end{equation}
$RPV$ couplings $\lambda'_{ijk}$ induce tree
level contributions via squark exchanges to the decay $K^+\to
\pi^+\nu\bar{\nu}$ (cf the diagrams shown in Fig.~\ref{fig1}), so 
only the second term will be considered here.
%%%%%%%%%%%%%%
%%%%%%%%%%%%%%

The branching ratio for 
the rare decay $K^+ \to \pi^+ \nu \bar{\nu}$ can be written in the 
following form:
\begin{equation} 
\label{KSM}
BR(K^+ \to \pi^+ \nu \bar{\nu})\propto
\left( \sum_l \mid C_l^{RPC} + \epsilon_{ll} \mid^2 + \sum_{k\not= l} \mid \epsilon_{kl} \mid^2 \right) 
\end{equation}
where $C_l^{RPC}$ is the $R$-parity conserving contribution proportional
to ($\lambda_cX_c^l + \lambda_t\,X_\mathit{new}$), the sum is over $\nu$'s and
anti-$\nu$'s flavours, and the $R$-parity violating 
couplings are contained in the $\epsilon_{ij}$ :
%%%%%%%%%%%%%%
\begin{equation} 
\epsilon_{ij}= \sum_n \left( \frac{\lambda_{i2n}^{'*}\lambda_{j1n}^{'}}
{m_{\tilde{d}n_R}^2} -\frac{\lambda_{in1}^{'*}\lambda_{jn2}^{'}}
{m_{\tilde{d}n_L}^2} \right)\; .
\end{equation}
%%%%%%%%%%%%%%
Under some assumption it is possible to
constrain certain combinations of couplings, which will be done in the next
section.

%%%%%%%%%%%%%%%%%%%%%%%%%%%%%%%%%%%%%%%%%%%%%%%%%%%%%%%%%%%%%%%%%%%%%%%%%%%%%%%%
%%%%%%%%%%%%%%%%%%%%%%%%%%%%%%%%%%%%%%%%%%%%%%%%%%%%%%%%%%%%%%%%%%%%%%%%%%%%%%%%
\section{Constraints}%%%%%%%%%%%%%%%%%%%%%%%%%%%%%%%%%%%%%%%%%%%%%%%%%%%%%%%%%%%
%%%%%%%%%%%%%%%%%%%%%%%%%%%%%%%%%%%%%%%%%%%%%%%%%%%%%%%%%%%%%%%%%%%%%%%%%%%%%%%%
%%%%%%%%%%%%%%%%%%%%%%%%%%%%%%%%%%%%%%%%%%%%%%%%%%%%%%%%%%%%%%%%%%%%%%%%%%%%%%%%
From our previous discussion, we have drawn the conclusion that $RPC$
supersymmetry 
has to be included in the analysis of $K^+ \to \pi^+ \nu \bar{\nu}$.
Since we aim to obtain an upper-bound on the $RPV$ couplings, we assume the 
$RPC$ contributions (which already include $SM$ ones)
to be minimal (corresponding to $r_K = 0.5$ and $\theta_K =
25^o$) in order to allow for the largest possible contribution from $RPV$ terms.

In contrast to the standard model and the RPC supersymmetric contributions, 
$R$-parity violating couplings can induce tree level processes with 
a neutrino and an antineutrino of different flavour in the final state.
The $R$-parity violating processes with the same neutrino flavour in the 
final state, then,
interfere with the $SM/RPC$ $SUSY$ contributions as can be seen from the first
term in Eq.~\ref{KSM}. 

Just to see later the effect of the interferences, we neglect them in a first
step. This leads to the bounds (setting all the couplings to zero except one 
product) :
%%%%%%%%%%%%%% 
\begin{equation} 
\mid\frac{\lambda_{i2n}^{'*}\lambda_{i1n}^{'}}
{m_{\tilde{d}n_R}^2} \mid\ ,\ \mid\frac{\lambda_{in1}^{'*}\lambda_{in2}^{'}}
{m_{\tilde{d}n_L}^2} \mid\ < \frac{2.1\times 10^{-5}}{(200\ \mathrm{GeV})^2}
\label{int-limits1}
\end{equation}
%%%%%%%%%%%%%%

\begin{figure}[htb]
\begin{center}
  \mbox{\epsfxsize=0.4\textwidth
       \epsffile{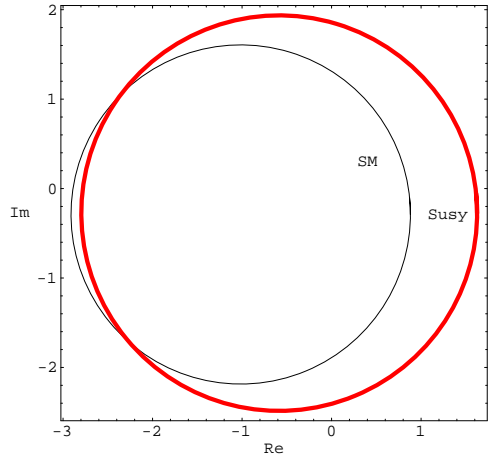}}
  \end{center}
\caption[]{\label{fig:plot1sigma} Allowed region for
$\mathrm{Re}(\epsilon_{11})$ and $\mathrm{Im}(\epsilon_{11})$ in units of
$10^{-5}$, for the case of the standard model ($r_K = 1$ and
$\theta_K = 0$, thin black circle) and for the ``minimal'' $RPC$ $SUSY$ ($r_K = 0.5$ 
and $\theta_K = 25^o$, red circle). The reference value for the mass of the 
squarks is 200 GeV.}
\end{figure}

But more realistic and precise constraints should take into
account interferences. This, however, makes the extraction of upper
bounds harder and no simple bounds can be given. In the following, 
we will assume that only final states with the same neutrino flavour occur. 
Thus, only $\epsilon_{ij}$ with $i = j$ has to be taken
into account. The general equation verified by the $\epsilon_{ii}$ can
be written in the following way:
%%%%%%%%%%%%%%
\begin{equation} 
\sum_{i=e,\mu,\tau}\left(\mathrm{Re}(\epsilon_{ii}) + \frac{\alpha_i}{2}\right)^2\; +  
\sum_{i=e,\mu,\tau}\left(\mathrm{Im}(\epsilon_{ii}) + \frac{\beta}{2}\right)^2=R^2\; .
\label{cercleeq}
\end{equation}
%%%%%%%%%%%%%%
Taking only one of the $\epsilon_{ii}$ nonzero, this equation
describes a circle in the complex plane, whose parameters can be found in the
original paper~\cite{DOW}. As an example, the resulting constraints in the complex plane 
on $\epsilon_{11}$ are displayed in
Fig.~\ref{fig:plot1sigma}. 
To have a numerical idea of the interferences, we may choose the point
of coordinates (Re($\epsilon_{11}$)=-2, Im($\epsilon_{11}$)=-2) on the ``$SUSY$"
circle of Fig.2. It is approximately the point which gives the maximum value
for $\mid \epsilon_{11} \mid$ : $\mid \epsilon_{11} \mid_{max}\simeq2.8\times
10^{-5}$. That leads to :
\begin{equation} 
\mid\frac{\lambda_{i2n}^{'*}\lambda_{i1n}^{'}}
{m_{\tilde{d}n_R}^2} \mid\ ,\ \mid\frac{\lambda_{in1}^{'*}\lambda_{in2}^{'}}
{m_{\tilde{d}n_L}^2} \mid\ < \frac{2.8\times 10^{-5}}{(200\ \mathrm{GeV})^2}
\label{int-limits2}
\end{equation}
Constraints on $\epsilon_{22}$ and
$\epsilon_{33}$ can be obtained in the same way and are of the same
order of magnitude, the limits~(\ref{int-limits2}) can be used for the 3 flavours.  

These upper-bounds are 30$\%$ bigger than without interferences and so, 
our conclusion is that interferences do have a significant influence. 

%%%%%%%%%%%%%%%%%%%%%%%%%%%%%%%%%%%%%%%%%%%%%%%%%%%%%%%%%%%%%%%%%%%%%%%%%%%%%%%%
%%%%%%%%%%%%%%%%%%%%%%%%%%%%%%%%%%%%%%%%%%%%%%%%%%%%%%%%%%%%%%%%%%%%%%%%%%%%%%%%
\section*{References}%%%%%%%%%%%%%%%%%%%%%%%%%%%%%%%%%%%%%%%%%%%%%%%%%%%%%%%%%%%
%%%%%%%%%%%%%%%%%%%%%%%%%%%%%%%%%%%%%%%%%%%%%%%%%%%%%%%%%%%%%%%%%%%%%%%%%%%%%%%%
%%%%%%%%%%%%%%%%%%%%%%%%%%%%%%%%%%%%%%%%%%%%%%%%%%%%%%%%%%%%%%%%%%%%%%%%%%%%%%%%

\end{document}